\documentclass[conference]{IEEEtran}
\IEEEoverridecommandlockouts
\usepackage{cite}
\usepackage{amsmath,amssymb,amsfonts}
\usepackage{algorithmic}
\usepackage{graphicx}
\usepackage{textcomp}
\usepackage{xcolor}
\usepackage{url}
\usepackage[top=0.8in, bottom=1.05in, left=0.8in, right=0.8in, includefoot]{geometry}
\def\BibTeX{{\rm B\kern-.05em{\sc i\kern-.025em b}\kern-.08em
    T\kern-.1667em\lower.7ex\hbox{E}\kern-.125emX}}
\begin{document}

\title{A Decentralized Quantum-Resistant Notary Scheme for Blockchain Interoperability with a Novel Dynamic Trust Model}

\author{
  \IEEEauthorblockN{
    Koosha Esmaeilzadeh Khorasani\IEEEauthorrefmark{1},
    Saeed Moradi\IEEEauthorrefmark{1},
    Sara Rouhani\IEEEauthorrefmark{2}
  }
  \IEEEauthorblockA{\IEEEauthorrefmark{1}Department of Computer Science, University of Manitoba, Winnipeg, Canada \\
  \IEEEauthorrefmark{2}University of Calgary, Calgary, Canada \\
  Email: \{esmaeilk, moradis1\}@myumanitoba.ca, sara.rouhani@ucalgary.ca}
}

\maketitle

\begin{abstract}
Blockchain interoperability is essential for enabling secure and reliable interactions across heterogeneous networks. However, existing notary schemes face two key issues: quantum vulnerability and rigid trust management. To address the first issue, this paper proposes a decentralized, quantum-resistant framework integrating post-quantum primitives ML-KEM and ML-DSA alongside AES-256, optimizing performance by using asymmetric cryptography for authentication and non-repudiation, and delegating bulk encryption to symmetric schemes. Under parallel loads of up to 270 transactions per second, this hybrid design reduces the end-to-end performance gap between the post-quantum and elliptic-curve approaches to under 6\% on average, converging to within 0.2\% at peak load. The second issue is handled by a dynamic trust model that uses Popoviciu’s inequality to derive an adaptive discount factor $\alpha$  based on variance in a Verifier's behavior, removing the need for fixed thresholds or multi-stage configuration, while a tunable dynamicity factor $k$ enables control over system sensitivity. We evaluated the trust model over 200 epochs under adversarial conditions, where the proposed model detects and responds to misbehavior that the baseline model fails to penalize adequately. These results demonstrate that quantum-resistant blockchain interoperability is practically achievable, with the proposed trust mechanism offering greater adaptability and resilience than existing approaches.

\end{abstract}


\section{Introduction}
\label{sec:introduction}
Blockchain is a distributed system that enables decentralized governance 
through diverse and heterogeneous networks, such as Ethereum 
\cite{buterin2013ethereum} and Monero \cite{alonso2020zero}. Differences in blockchain protocols and architectures prevent direct 
cross-chain communication, necessitating dedicated interoperability 
mechanisms \cite{hassan2026sok}, Moreover, most interoperability solutions assume that networks are already
mutually aware, whereas the initial discovery of blockchain networks and their services requires its own decentralized mechanism~\cite{hassan2025enabling}.
Existing approaches include Blockchain of Blockchains, Sidechains,  Trusted Relays, Hashed Time-Lock Contracts, and Notary Schemes 
\cite{ren2023interoperability,belchior2021survey}. However, these 
mechanisms introduce security risks that can result in significant  financial losses \cite{10646648}, making the security and integrity of  cross-chain interactions fundamental requirements for blockchain 
interoperability.

In this paper, we focus on two fundamental security challenges in blockchain interoperability: quantum resilience and adaptive trust management within notary-based protocols. Notary schemes are particularly relevant due to their prevalence and their clear and flexible trust structure, where Verifier nodes are responsible for validating cross-chain interactions.

From a confidentiality perspective, most existing interoperability solutions rely on classical cryptographic primitives such as Elliptic Curve Cryptography (ECC) \cite{koblitz1987elliptic} and RSA \cite{rivest1978method}. While secure against conventional attacks, these schemes are vulnerable to quantum adversaries, as demonstrated by Shor’s and Grover’s algorithms \cite{shor1999polynomial,grover1996fast}. Although the quantum threat has been acknowledged in blockchain research \cite{holcomb2021pqfabric,sharma2021analysis}, only limited efforts have been made to incorporate post-quantum cryptography into interoperability protocols \cite{yi2023post,wang2022secure}. To the best of our knowledge, no existing interoperability protocol integrates the newly standardized NIST post-quantum primitives, such as ML-KEM \cite{fips2024203} and ML-DSA \cite{fips2024204}, within a complete cross-chain validation pipeline.

The second major challenge lies in trust management. In notary-based systems, notaries play a critical role in validating cross-chain operations, yet existing trust mechanisms are often static or weakly adaptive. Many approaches rely on fixed weighting schemes \cite{cao2023cross, chen2024cross}, leading to limited responsiveness under dynamic or adversarial conditions. Other methods adjust trust scores but follow rigid update patterns, resulting in either abrupt or overly slow reactions to malicious behavior \cite{putra2023privacy, putra2021trust}. More recent work \cite{hu2024blockchain} introduces dynamic evaluation mechanisms; however, it relies on multiple thresholds and lacks explicit support for adaptive trust recovery, limiting its practical responsiveness.

Therefore, existing interoperability solutions exhibit two fundamental limitations, namely reliance on classical cryptographic primitives that are not quantum-resistant and the use of static or weakly adaptive trust models that do not dynamically influence notaries' participation. More importantly, current designs treat trust as an external component, rather than integrating it into the verification process itself.

In order to address the aforementioned challenges, this research makes the following key contributions: i) a trust-adaptive, quantum-resistant interoperability framework that uses trust-based notary selection and lightweight quorum verification for efficient cross-chain validation; ii) integration of NIST-standardized ML-KEM and ML-DSA with AES-256 to provide post-quantum security with low cryptographic overhead; iii) an adaptive trust model based on Popoviciu’s inequality that balances responsiveness and stability through a tunable parameter $k$ to detect intermittent attacks; and iv) a prototype implementation and experimental evaluation demonstrating efficient validation, low overhead, and dynamic isolation of unreliable verifiers.
\section{Related Work}
\label{sec:RelatedWorks}

\subsection{Blockchain Interoperability}
Blockchain interoperability solutions can be broadly classified into \textit{architectural} and \textit{cryptographic approaches}. Architectural approaches rely on intermediary infrastructure and include several categories: (i) \textit{Blockchain of Blockchains}, using hub-based architectures such as Polkadot~\cite{wood2016polkadot} and Axelar~\cite{axelar2021}; (ii) \textit{Hashed Time-Lock Contracts (HTLCs)}, enabling atomic cross-chain transactions through hash-locks and time-locks, as in OTEx~\cite{dangat2024otex}; (iii) \textit{Relay schemes}, transferring and verifying cross-chain data, as in Xclaim~\cite{zamyatin2019xclaim} and Ivycross~\cite{li2021ivycross}; (iv) \textit{Sidechains}, supporting cross-chain asset and data transfers, such as Loom Network~\cite{loomxio} and Liquid~\cite{liquid2020}; and (v) \textit{Notary schemes}, where trusted entities or committees verify cross-chain transactions, as in Bifröst~\cite{scheid2019bifrost} and BAILIF~\cite{karumba2023bailif}.

Privacy-preserving notary schemes have incorporated ECC, AES, BLS, and ERC20-based mechanisms~\cite{tan2024study,bacho2022adaptive,victor2019measuring}. Wu et al.~\cite{wu2023distributed} employed group signatures and blind identities to preserve notary privacy while enabling regulatory accountability. Post-quantum approaches include multivariate cryptography-based multi-signatures~\cite{yi2023post} and identity-based quantum multi-signature protocols with Quantum Freeze mechanisms~\cite{wang2022secure}.

Cryptographic interoperability approaches reduce reliance on intermediaries through specialized primitives. \textit{Adaptor signatures} enable atomic cross-chain swaps using conditional signatures~\cite{gerhart2024foundations}, while cryptographic oracle-based conditional payments~\cite{madathil2022cryptographic} and verifiable witness encryption~\cite{dottling2023mcfly} enforce transaction conditions cryptographically. However, these mechanisms may require specialized transaction functionality or additional protocol complexity.

\subsection{Trust and Notary Scheme}

Trust management in blockchain solutions has been applied across various domains, including the Internet of Things (IoT) \cite{putra2023privacy}, supply chain management \cite{batwa2021blockchain}, and vehicular networks \cite{yang2018blockchain}, among others. However, as our proposed solution specifically addresses blockchain interoperability using Notary Schemes, this will be the primary focus of our discussion.

Trust evaluation in notary schemes has been addressed through various approaches. Chen et al.~\cite{chen2024cross} combined PageRank~\cite{page1999pagerank} with entropy weighting to assess notary trust using transaction history, node similarity, and trust relationships. Guo and Hu~\cite{guo2024calculation} employed a modified EigenTrust algorithm for reputation evaluation, integrating verifiable random functions (VRFs) and cumulative probability for notary selection. Cao and Yang~\cite{cao2023two} enhanced PageRank with node quality, initial scores, and age-related factors, combining VRFs with a half-life mechanism to reduce reputation bias toward older nodes. Cao et al.~\cite{cao2023cross} proposed a cross-chain traceability framework involving authorization, access control, and notary management, where reputation is determined by voting outcomes, transaction efficiency, and feedback.

\begin{figure}
	\centering
	\includegraphics[width=\columnwidth,height=5cm]{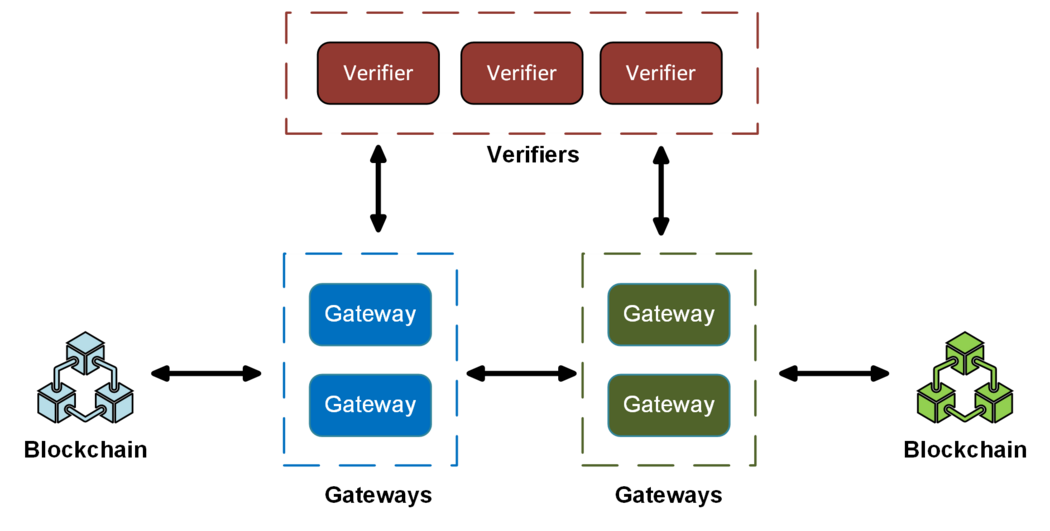}
	\caption{Overview of the proposed system architecture.}
	\label{FIG:1}
\end{figure}
Taking a more dynamic approach, Hu et al. \cite{hu2024blockchain} developed an interoperability solution featuring a dynamic reputation system. Their model evaluates notary nodes using network trust and committee assessments over time. It introduces a penalty factor for malicious actions, a dynamic evaluation window algorithm leveraging a recurrent neural network (RNN) to adapt intervals based on network conditions, and a reputation decay mechanism to address inactivity.

\section{System and Trust Modeling}

To tackle the challenges outlined earlier, we present a notary-based interoperability solution for heterogeneous blockchains, leveraging our proposed trust model and post-quantum cryptography. The notary scheme was selected for its scalability, flexibility, and ability to allow network participants to act as notaries. The system consists of two main components: Gateways, which connect a blockchain to the interoperability network and act as entry and exit points, and Verifiers, which serve as notaries. To join the permissioned interoperability network, each participant must provide at least one Gateway and one Verifier. Figure \ref{FIG:1} illustrates the relationship between these components.
\begin{table}[htbp]
\caption{Notation Table}
\begin{center}
\begin{tabular}{|c|c|}
\hline
\textbf{Symbol} & \textbf{Description} \\
\hline
$B_i$ & Blockchain $i$ \\
\hline
$g_i$ & Gateway $i$ \\
\hline
$pk$ & Public key \\
\hline
$sk$ & Secret key \\
\hline
$v_i$ & Verifier $i$ \\
\hline
$\text{sym\_key}$ & Symmetric key \\
\hline
$\text{addr}_{\text{n}}$ & Address of a node \\
\hline
$\mathcal{TS}$ & Trust score \\
\hline
$\sigma$ & Digital signature \\
\hline
$\text{Msg}_{\text{kem}}$ & Key encapsulation request message \\
\hline
$\text{ct}$ & Ciphertext generated in key encapsulation \\
\hline
$\text{Msg}_{\text{sync}_{g_i}}$ & Synchronization message introducing $g_i$ to Verifiers \\
\hline
$\text{Msg}_{\text{GetInfo}}$ & Message for receiving network information \\
\hline
$\textit{trans}_{\textit{req}}$ & Request for state transition \\
\hline
$msg_{\text{transition}}$ & Message containing $\textit{trans}_{\textit{req}}$ \\
\hline
$V_{cc}$ & Verifier consensus committee \\
\hline
\end{tabular}
\label{tbl:notation}
\end{center}
\end{table}

We model the system in three phases. In \textbf{Initialization}, nodes perform key generation, key encapsulation, and information dissemination upon enrollment. In \textbf{Cross-chain Interaction}, tickets are issued and messages authenticated via a Kerberos-inspired protocol~\cite{neuman1994kerberos}, subject to validation at the destination Gateway. Finally, \textbf{Consensus and Trust Management} comprises the consensus procedure initiated by Gateways and the trust adjustment mechanism among Verifiers.

\subsection{Initialization}

\subsubsection{Key Generation}
Each node (Gateway or Verifier) generates the required cryptographic key pairs upon launch. These include a key pair for digital signatures $(pk_{\text{sig}}, sk_{\text{sig}})$; a key pair for key encapsulation used for encryption $(pk_{\text{kem}}^{\text{enc}}, sk_{\text{kem}}^{\text{enc}})$; a second key encapsulation pair used for HMAC operations $(pk_{\text{kem}}^{\text{hmac}}, sk_{\text{kem}}^{\text{hmac}})$; the address and public key of the bootstrap node $(\text{addr}_{\text{b}}, pk_{\text{b}})$; and two symmetric keys derived from key encapsulation, one for encryption ($\text{sym\_key}^{\text{enc}}$) and one for HMAC-based message authentication ($\text{sym\_key}^{\text{hmac}}$). In addition, each Gateway holds a reference $B_i$ to the blockchain with which it directly interacts, while each Verifier maintains a trust score $\mathcal{TS}_{v_i}$ determined by its participation in the consensus.

\subsubsection{Key Encapsulation}
Each node performs key encapsulation with a Bootstrap Verifier $v_B$, whose address and public key are assumed to be known in advance. To establish a shared symmetric key with $v_B$, a Gateway $g_i$ sends a request message
\begin{align*}
\text{Msg}_{\text{kem}} = (& pk_{\text{kem}}, \text{addr}_g, \\
                          & \text{dsa\_scheme\_name},\\ & \text{kem\_scheme\_name})
\end{align*}
accompanied by its signature, defined as $\sigma = \text{ML-DSA.Sign}(\text{Msg}_{\text{kem}}, sk_{\text{sig}})$, where the scheme-name fields specify the ML-DSA and ML-KEM versions in use. Upon receipt, $v_B$ verifies the signature via $\text{ML-DSA.Verify}(\text{Msg}_{\text{kem}}, \sigma, pk_{\text{sig}})$ and, if successful, stores the Gateway's public keys and performs encapsulation:
\begin{align*}
\text{sym\_key}_{v,g_i}^{\text{enc}}, \text{ct} = \text{ML-KEM.Encaps}(pk_{\text{kem}}).
\end{align*}
The ciphertext is signed and returned to $g_i$, which regenerates the shared key as $\text{sym\_key}_{v,g_i}^{\text{enc}} = \text{ML-KEM.Decaps}(sk_{\text{kem}}, ct)$. To enable secure communication between $g_i$ and the Verifier network, $v_B$ also distributes the shared key and associated metadata to the other Verifiers via a synchronization message:
\begin{align*}
\text{Msg}_{\text{sync}_{g_i}}
&= (\text{sym\_key}_{v,g_i}^{\text{enc}}, pk_{\text{sig}}, pk_{\text{kem}}, \sigma, \\
&\quad \text{addr}_g, \text{dsa\_scheme\_name}, \\
&\quad \text{kem\_scheme\_name})
\end{align*}
This procedure is repeated at fixed time intervals, and is likewise applied whenever new Verifiers join the system and must establish trust with $v_B$.

\subsubsection{Get Information}
After key distribution, a node requests network information from $v_B$. Since non-repudiation is not critical here, the request is authenticated with an HMAC (SHA-256, keyed with a symmetric key established via the same procedure described above) rather than a digital signature:
\begin{align*}
\text{Msg}_{\text{GetInfo}_{g_i}} &= (\text{req\_id}, \text{HMAC}_{\text{sym\_key}_{v,g_i}^{\text{hmac}}}(\text{req\_id}))
\end{align*}
In response, $v_B$ returns all the data it maintains about the Gateways and Verifiers in the system---IP addresses, ports, public keys, and other metadata---for use in subsequent communication.

\subsection{Cross-chain Interaction}
To enable cross-chain interaction between blockchains $B_i$ and $B_j$, their Gateways $g_i$ and $g_j$ communicate through a Kerberos-inspired mechanism. First, $g_i$ requests a ticket from a Verifier, which generates a 32-byte symmetric session key $\text{session\_key}_{g_{ij}}^{\text{enc}}$ using PBKDF2 and SHA-256  to secure Gateway-to-Gateway communication.
The Verifier then constructs a ticket
\begin{align*}
\text{Ticket} = (\text{session\_key}_{g_{ij}}^{\text{enc}}, pk_{g_i}^{\text{sig}}, addr_{g_i}, lifetime)
\end{align*}
where \textit{lifetime} defines the ticket's validity period. The ticket is encrypted under $\text{sym\_key}_{g_j}^{\text{enc}}$, the key previously established with $g_j$, and returned to $g_i$ together with $\text{session\_key}_{g_{ij}}^{\text{enc}}$ encrypted under $\text{sym\_key}_{g_i}^{\text{enc}}$.

Gateway $g_i$ then constructs a transition request $\textit{trans}_{\textit{req}}$ containing the operation details, encrypts it with the session key, and sends
\begin{align*}
msg_{\text{transition}} = (\textit{trans}_{\textit{req}}, \text{Ticket}, \sigma)
\end{align*}
to $g_j$. Upon receipt, $g_j$ decrypts the ticket to obtain the session key, decrypts $\textit{trans}_{\textit{req}}$, and verifies the signature using $g_i$'s public key extracted from the ticket. After validation, $g_j$ executes the state transition on $B_j$ and returns a response message
\begin{align*}
msg_{\text{response}} = (\textit{trans}_{\text{res}}, \text{Proof}),
\end{align*}
encrypted with $\text{session\_key}_{g_{ij}}^{\text{enc}}$, where $\textit{trans}_{\text{res}}$ is the transaction response and $\text{Proof}$ is generated using \textit{zk-SNARKs} \cite{petkus2019and}, specifically the \textit{Groth16} protocol \cite{groth2016size}, chosen for its efficiency and succinct proof size.


\subsection{Consensus Mechanism and Trust Management}
Upon receiving $msg_{\text{response}}$, Gateway $g_i$ extracts the proof and forwards it to the consensus committee,
\[
\mathcal{V}_{\text{cc}} = \{ v \in \mathcal{V} \mid \mathcal{TS}_{v} > T \},
\]
the set of Verifiers whose trust scores exceed the threshold $T$, where $\mathcal{V}$ is the complete set of Verifiers and contains at least $3f+1$ nodes. The threshold $T$ is selected such that $|\mathcal{V}_{\text{cc}}| \geq f+1$, ensuring that a quorum can be formed under the assumed fault model.

Each committee member independently verifies the proof, returns its result to $g_i$, and exchanges its validation outcome with the other members; this mutual exchange drives the dynamic adjustment of trust scores. Once $g_i$ collects at least $f+1$ consistent responses, it either applies the result on $B_i$ or initiates a rollback, which may include a compensating action on $B_j$ depending on the operation. Finally, depending on the protocol, $g_i$ may in turn generate a proof for $g_j$, verified through the same process.

\subsection{Trust Model}

In this section, we introduce our trust score calculation model, which extends the model of Putra et al.~\cite{putra2023privacy}:
 
\begin{equation}
\mathcal{TS}(t) = (1-\alpha) \sum_{i=1}^{t} \phi_t^{v_i} \alpha^{(t-i)},\,
 \quad 0 < \alpha < 1\,,
\end{equation}
 
where $\mathcal{TS}$ denotes the trust score, $\alpha$ the discount factor, and $\phi$ whether the node acted honestly or maliciously during proof verification. While this formulation emphasizes recent activity, it does so at a fixed rate. In practice, it is beneficial to consider the entire behavioral history when Verifiers produce consistent results, but essential to weight recent activity more heavily when anomalies arise. We therefore make the trust score \textit{dynamic}, using the variance of the verification results ($\phi_t^{v_i}$) as an indicator: consistent alignment with consensus (zero variance) signals a stable Verifier, whereas inconsistent outcomes are treated as anomalies that shift the emphasis toward recent activity.
 
When the variance is zero, we assign the most recent verification result, $\phi_t^{v_i}$, as the trust score of Verifier $v_i$ in epoch $t$, where each epoch corresponds to the completion of one consensus process. Although assigning high initial trust may appear optimistic, it is justified in private interoperability networks, where nodes are semi-trusted and added by known participants; as shown in subsequent sections, the model sharply reduces a node's trust score upon even a small number of malicious actions. If the results exhibit inconsistency (non-zero variance), we instead employ the exponentially discounted score:
 
\begin{equation}
\label{eq_trust}
\mathcal{TS}_{v_i}(t) =
\begin{cases}
\phi_t^{v_i}, & \text{if Var} = 0, \\[6pt]
(1-\alpha) \sum_{j=1}^{t} \phi_j^{v_i}\,\alpha^{(t-j)},
& \begin{aligned}[t]
\text{if Var} > 0, \\
 0<\alpha<1
\end{aligned}
\end{cases}
\end{equation}
where
\begin{equation}
\phi_t^{v_i} =
\begin{cases}
1, & \text{if Verifier $v_i$ acts benevolently}\,, \\[6pt]
0, & \text{if Verifier $v_i$ acts maliciously}\,.
\end{cases}
\end{equation}
 
Rather than a fixed discount factor, we compute $\alpha$ dynamically from the variance of the verification history:
 
\begin{equation}
\label{eq3}
\alpha = \exp\left(-k \cdot \frac{\mathrm{Var}}{\mathrm{Var}_{\max}}\right),
\end{equation}
where $\mathrm{Var}$ is the variance of $\phi^{v_i}$ and, by Popoviciu's inequality~\cite{egozcue2018variance}, $\mathrm{Var}_{\max} = (\phi_{\max} - \phi_{\min})^2/4 = 1/4$ since $\phi^{v_i} \in [0,1]$. The \textit{dynamicity factor} $k$ controls the steepness of trust adjustments: a larger $k$ amplifies the effect of variance, producing sharper transitions, while a smaller $k$ dampens it, yielding smoother transitions.

\section{Implementation and Evaluation}
\subsection{Implementation and System Configuration}
The proposed model was implemented\footnote{\url{https://github.com/tcdt-lab/PQ-NS-IOP}} and simulated in Go using modular components for scalability and maintainability. The integrated modules emulate interactions among participating entities to evaluate the architecture's correctness, feasibility, and performance. Experiments were conducted on a system with 64 processing units, 128~GB RAM, and 512~GB SSD, with two Gateways and five Verifiers deployed to simulate the proposed architecture.


\subsection{Evaluations}
We evaluate the performance overhead of post-quantum cryptography compared with elliptic-curve cryptography and assess the trust mechanism's effectiveness against malicious behavior and traditional trust models.

\subsubsection{Key Encapsulation Mechanisms Evaluation}
\begin{figure*}
\centering
\includegraphics[width=.85\textwidth, height=4.5cm] {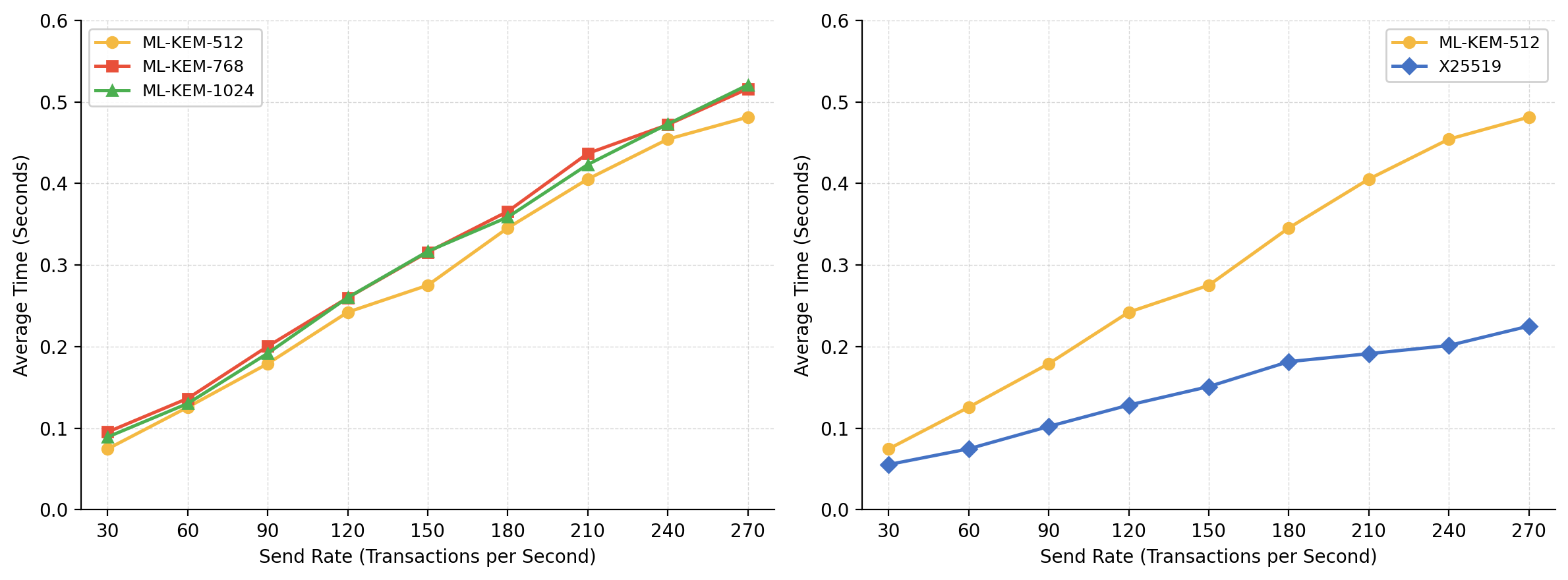}
\caption{(Left) Comparison of average key encapsulation time across different versions of ML-KEM. (Right) Performance comparison between ML-KEM-512 and X25519, highlighting the impact of underlying cryptographic foundations on efficiency.}
\label{FIG:7}
\end{figure*}

We evaluated KEM performance under parallel cryptographic operations, increasing the transaction rate from 30 to 270 TPS. Figure~\ref{FIG:7}~(Left) compares ML-KEM-512, ML-KEM-768, and ML-KEM-1024, showing that larger key sizes increase bandwidth overhead and degrade performance.

Figure~\ref{FIG:7}~(Right) compares ML-KEM-512 with X25519, which offers a comparable classical security level. X25519 significantly outperforms ML-KEM-512, highlighting the latency overhead of lattice-based quantum-resistant cryptography in the proposed notary scheme.

\subsubsection{Message Authentication and Integrity Evaluation}

\begin{figure*}
\centering
\includegraphics[width=.85\textwidth, height=4.5cm]{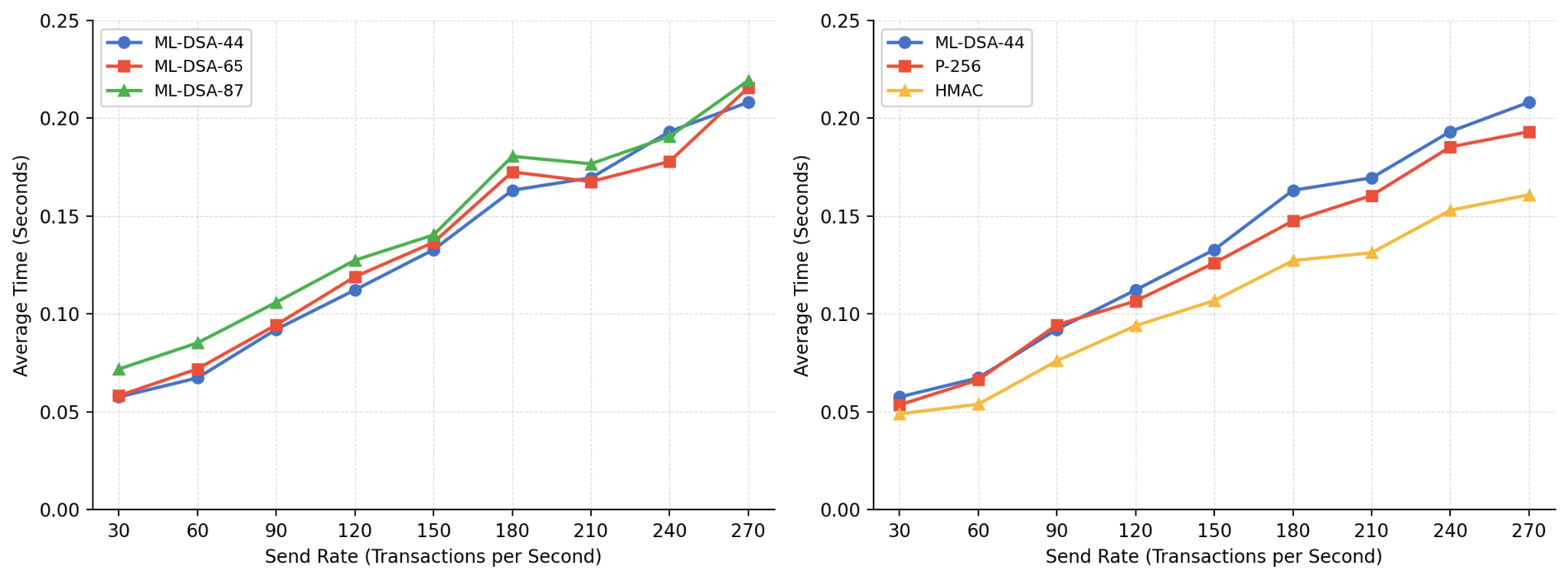}
\caption{Authentication performance under concurrency: (Left) shows the effect of key size in ML-DSA variants; (Right) compares ML-DSA-44, P-256, and HMAC-SHA256, highlighting HMAC’s efficiency.}
\label{FIG:9}
\end{figure*}
To evaluate the performance of the component responsible for message authentication and integrity, we assessed three approaches: post-quantum digital signatures, elliptic curve-based digital signatures, and HMAC, gradually scaling the load from 30 to 270 parallel transactions per second. We first compared the ML-DSA variants; as depicted in Figure~\ref{FIG:9}~(Left), processing time increases with higher transaction rates, and larger key sizes introduce additional overhead due to bandwidth constraints. However, the choice of cryptographic algorithm has an even more pronounced impact on performance, as illustrated in Figure~\ref{FIG:9}~(Right). In this comparison, ML-DSA-44, P-256, and HMAC-SHA256—all providing 128-bit security—were evaluated. HMAC, leveraging symmetric cryptography, significantly outperformed the others. While P-256 generally outperformed ML-DSA-44, the gap between them was modest compared to the substantial advantage of HMAC.

\subsubsection{End-to-End Evaluation}
\label{subsubsec:end_to_end}
We measured end-to-end interoperability latency, from ticket request and proof generation to verifier responses, under parallel loads of 30–270 TPS. On-chain execution time was excluded to isolate the interoperability layer's overhead, as it varies across blockchain platforms and configurations.

\begin{figure}
\centering
\includegraphics[width=\columnwidth, height=4.5cm]{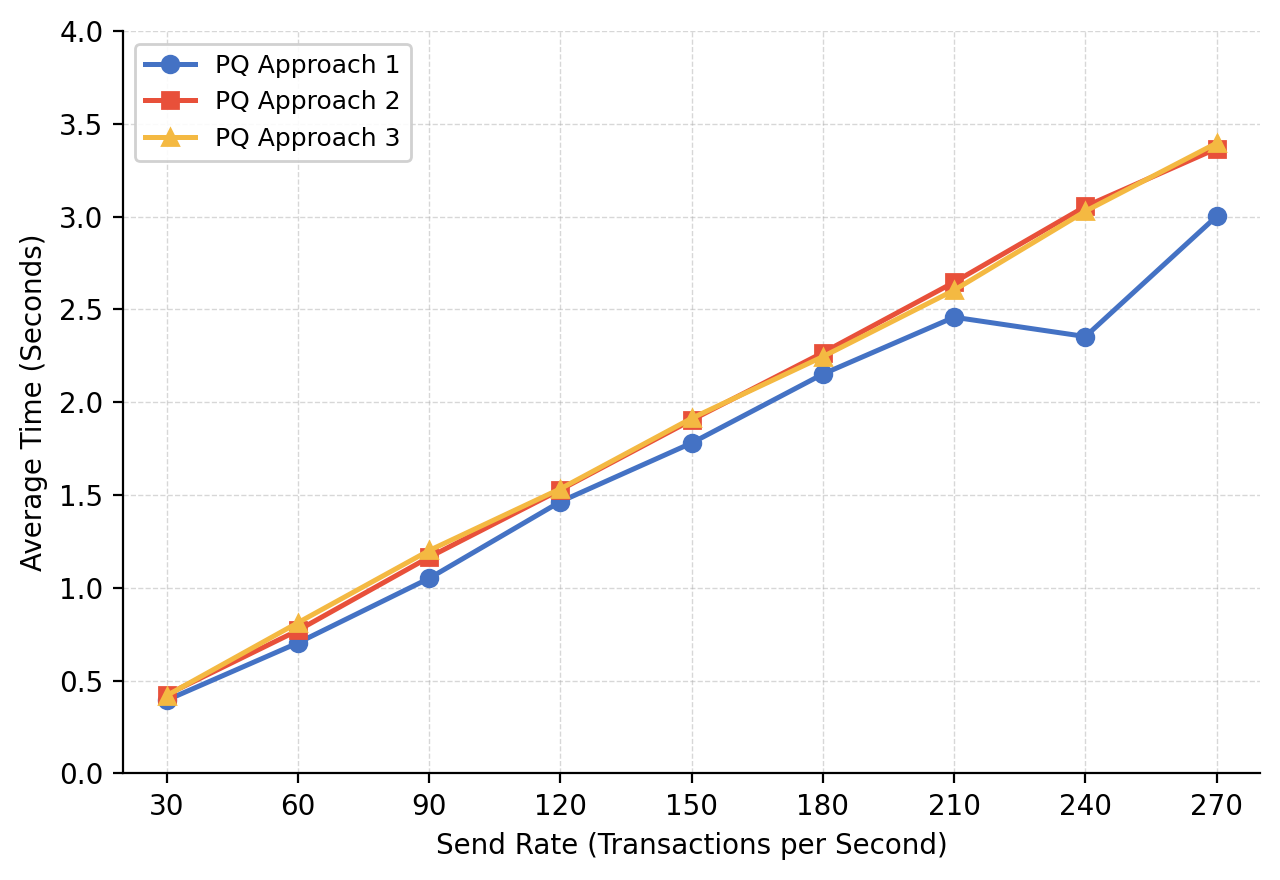}
\caption{Assessment of three post-quantum approaches according to NIST Security Categories: Approach 1 (Category II), Approach 2 (Category III), and Approach 3 (Category V).}
\label{FIG:10}
\end{figure}

\begin{figure}
\centering
\includegraphics[width=\columnwidth, height=4.5cm]{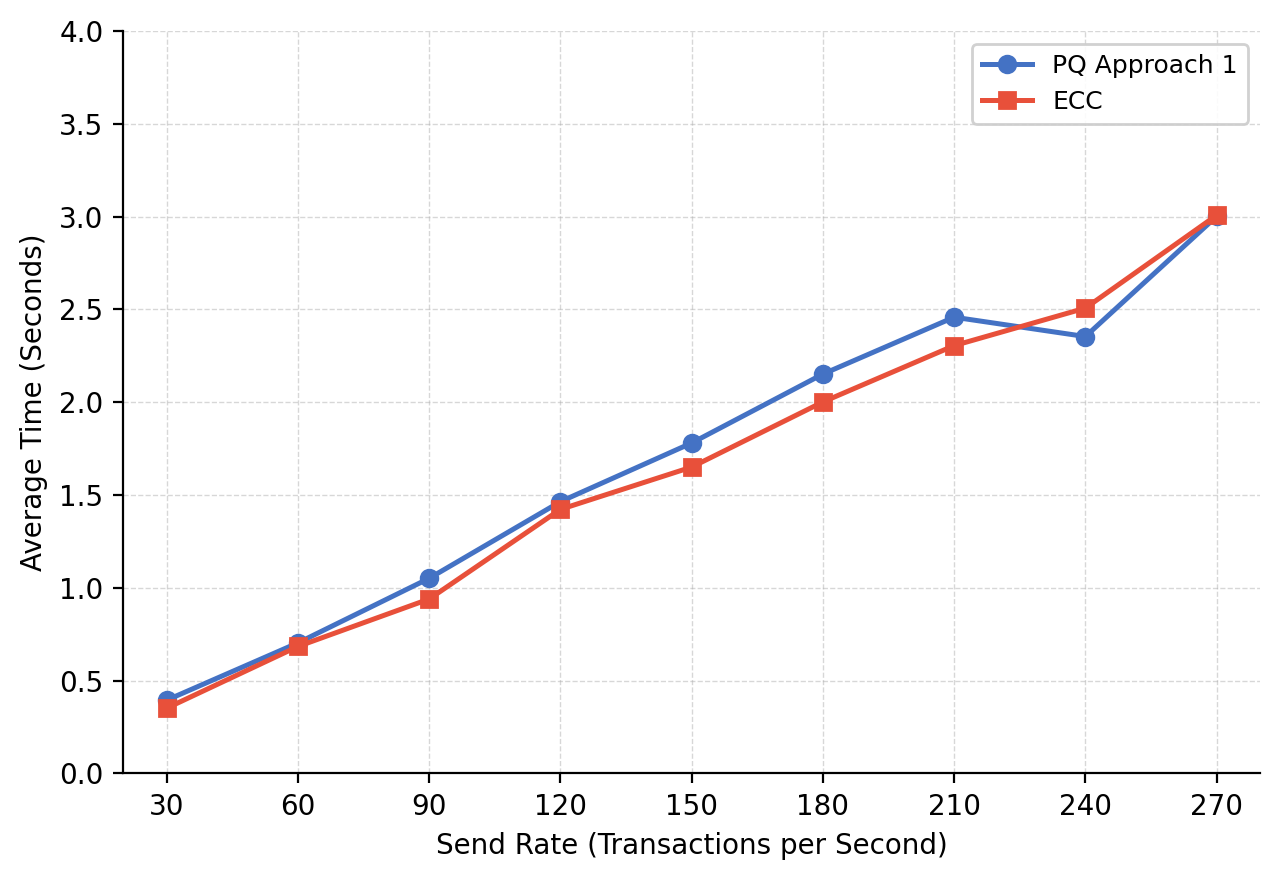}
\caption{Performance comparison between post-quantum Approach 1 and elliptic curve approach.}
\label{FIG:11}
\end{figure}

We first compared three post-quantum approaches based on ML-DSA-44 (Approach 1), ML-DSA-65 (Approach 2), ML-DSA-87 (Approach 3), corresponding to NIST Post-Quantum Security Categories II, III, and V respectively (Table \ref{table:sec_level}) to assess the effect of security level on end-to-end latency. As shown in Figure \ref{FIG:10}, the differences across categories remain within 10\% across all transaction rates, confirming that higher security levels introduce only marginal overhead in the proposed layered design.

We then evaluated a representative post-quantum configuration (Approach 1) against its elliptic-curve counterpart, using P-256 for digital signatures, in the end-to-end scenario. As shown in Figure \ref{FIG:11}, the post-quantum approach introduces an average overhead of approximately 5.1\% relative to the ECC baseline across the full load range, converging to within 0.2\% at 270 transactions per second. This narrow gap is a direct consequence of the architectural decision to confine asymmetric operations (whether post-quantum or classical) to authentication and non-repudiation tasks only, while delegating bulk message encryption to AES-256. Since asymmetric operations constitute a small fraction of total message processing time, substituting lattice-based primitives for elliptic-curve ones has limited impact on overall throughput. These results demonstrate that quantum-resistant interoperability is achievable at performance levels comparable to classical deployments under realistic parallel loads.

\subsubsection{Trust Model Evaluation}

This section presents the evaluation of the proposed trust model and the dynamic behavior of the discount factor in trust score calculation, which represents one of the key novelties of this research. The results are benchmarked against the baseline model introduced by Putra et al.~\cite{putra2023privacy}, which is based on the EigenTrust algorithm~\cite{kamvar2003eigentrust}.  

The dynamicity of the proposed trust model enables policy makers to design adaptive policies according to the specific requirements of the system. The policy adopted for our trust model follows a human-inspired approach, in which trust decreases sharply in the event of malicious behavior but is rebuilt gradually in response to consistent honest behavior. We implemented this policy by employing a dynamic $k$ in the trust model. Specifically, when incorrect validation results are detected, a higher value of $k$     $(k=15)$ is assigned, enabling a quicker response to inconsistencies. Conversely, when a Verifier consistently produces correct validation results, a lower value of $k$ $(k=0.09)$ is applied, reflecting the gradual process of rebuilding trust in a human-inspired approach.  
\begin{table}[htbp]
\caption{Security categories and their approximate strength equivalences.}
\begin{center}
\begin{tabular}{|c|c|}
\hline
\textbf{Security Category} & \textbf{Strength Equivalent} \\
\hline
I & At least as hard to break as AES-128 \\
\hline
II & At least as hard to break as SHA-256 \\
\hline
III & At least as hard to break as AES-192 \\
\hline
IV & At least as hard to break as SHA-384 \\
\hline
V & At least as hard to break as AES-256 \\
\hline
\end{tabular}
\label{table:sec_level}
\end{center}
\end{table}
Additionally, to demonstrate the flexibility of our model, we evaluated it at different values of $k$ to show how variations in this parameter influence the behavior of the trust score and discount factor. This highlights that a policy maker can adjust $k$ to adopt different trust policies or even combine multiple strategies, as we did in our implementation, to achieve tailored trust management policies.

\begin{figure*}
\centering
\includegraphics[width=.85\textwidth, height=4.5cm]{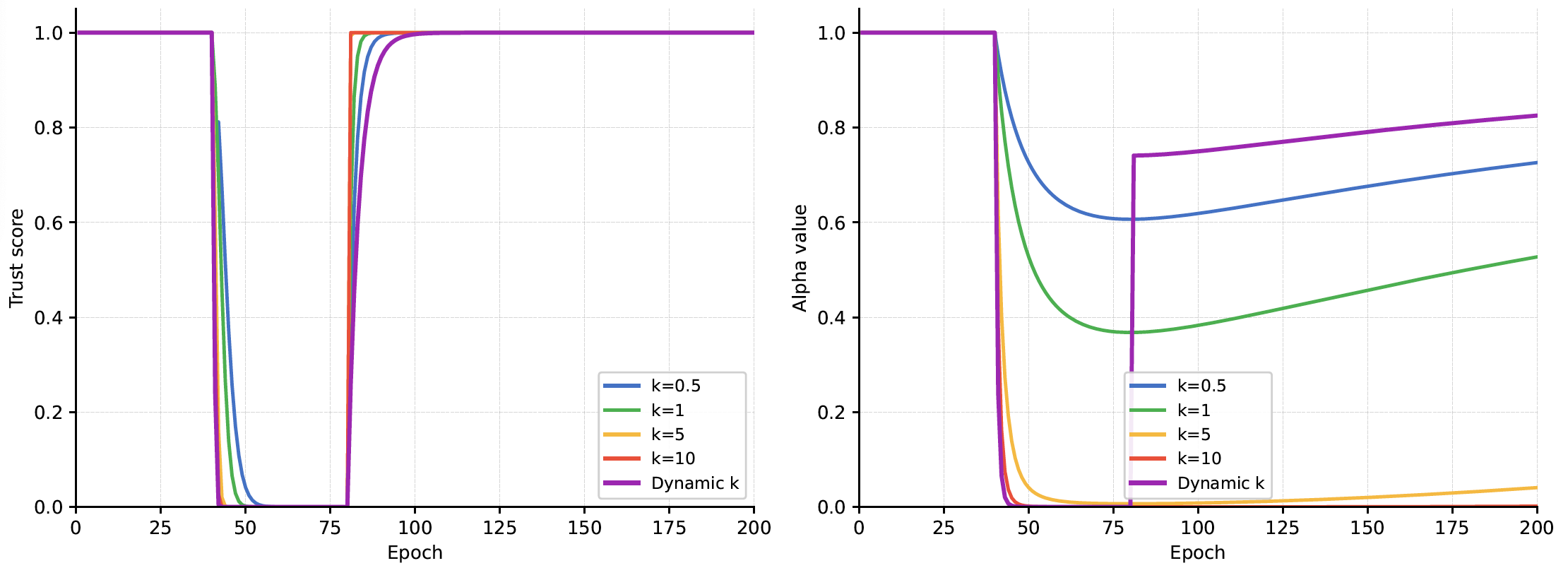}
\caption{Variation of trust score (Left) and discount factor (Right) across different values of \( k \) over epochs.}
\label{FIG:12}
\end{figure*}

The evaluation is conducted over 200 epochs, with incorrect transaction verifications deliberately introduced between epochs 40 and 80 to assess the model’s response under adversarial conditions. As illustrated in Figure~\ref{FIG:12}, increasing the value of $k$ from 0.5 to 10 results in more pronounced variations in the rate of trust score changes. Specifically, higher values of $k$ lead to steeper rates of both trust degradation and recovery. The dynamic $k$ illustrates the desired behavior of the model: a sharp decline in trust in response to malicious activity, followed by a more gradual recovery once the previously malicious node resumes honest behavior.  

With respect to the discount factor $\alpha$, higher values of $k$ cause a sharper decrease in $\alpha$, thereby emphasizing recent activities—a property particularly beneficial during periods of inconsistency. Conversely, lower values of $k$ give greater weight to historical data, stabilizing the trust score as honest behavior is re-established.  
\begin{figure}
\centering
\includegraphics[width=\columnwidth, height=4.5cm]{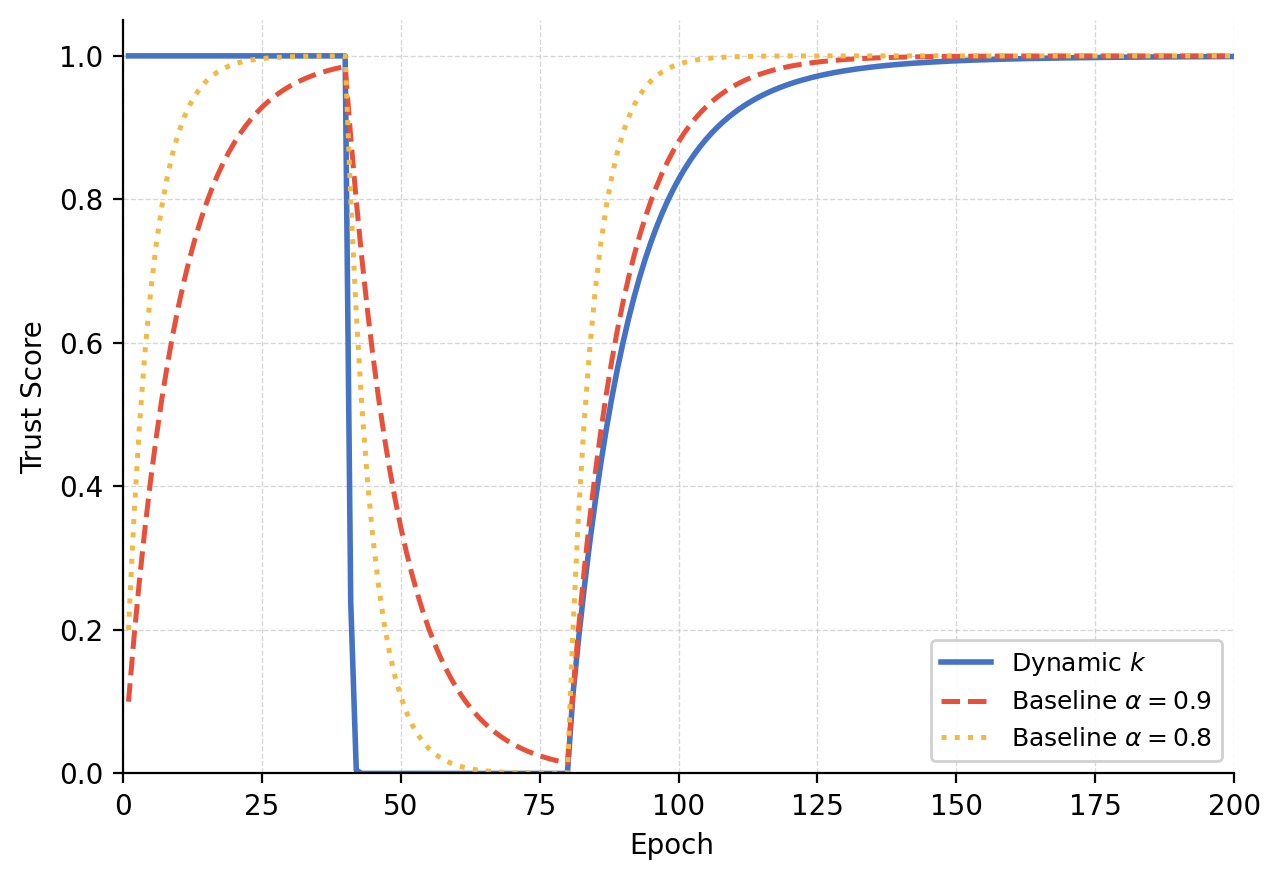}
\caption{Comparison of the baseline model with fixed \( \alpha \) values (\( \alpha = 0.8 \) and \( \alpha = 0.9 \)) against the dynamic \( k \) setting.}
\label{FIG:13}
\end{figure}

Figure~\ref{FIG:13} compares the proposed dynamic trust model with the fixed discount-factor baselines ($\alpha=0.8$ and $0.9$) from Putra et al.~\cite{putra2023privacy}. The proposed model offers three advantages. First, it incorporates initial trust in Verifiers, consistent with permissioned environments, and enables early threshold calibration based on the variance of $\phi_{v_i}$. Second, it penalizes malicious behavior more sharply while allowing gradual trust recovery. Finally, its dynamic parameters enable adaptive trust degradation and recovery according to system requirements.

\begin{figure*}
\centering
\includegraphics[width=.83\textwidth, height=4.5cm]{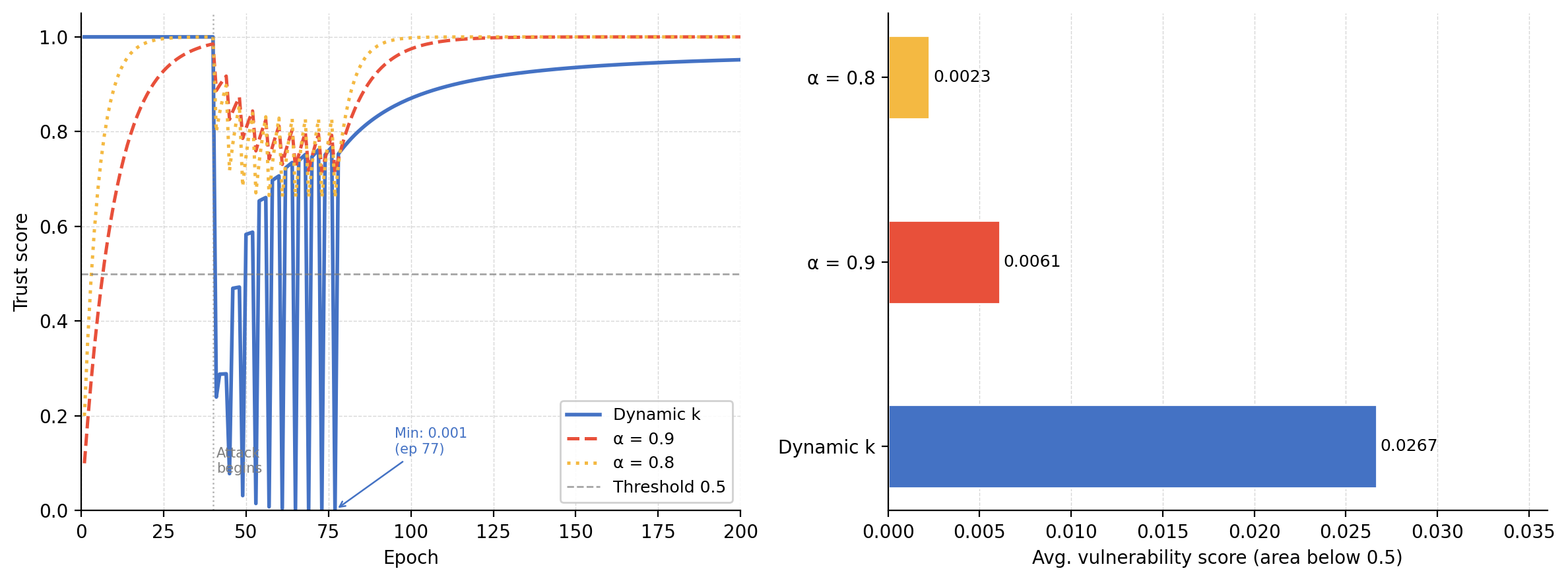}
\caption{Comparison of the baseline model with fixed $\alpha$ values ($\alpha = 0.8$ and $\alpha = 0.9$) and the dynamic $k$ setting in the proposed solution, where $v_i$ exhibits malicious behavior every four epochs between epochs 40 and 80. (Left) Trust score evolution over 200 epochs. (Right) Vulnerability score per strategy, computed as the mean area of trust score below the 0.5 threshold; higher values correspond to increased sensitivity.}

\label{FIG:14}
\end{figure*}

We further evaluate our proposed model in a scenario where a Verifier behaves maliciously once every four epochs between epochs 40 and 80. As illustrated in Figure~\ref{FIG:14}   (Left), the results show that the proposed model is more sensitive to sporadic malicious behavior. This heightened responsiveness enables the system to effectively detect Verifiers attempting to conceal dishonest actions among otherwise valid responses. This advantage stems from the dynamic $k$
mechanism: higher values of $k$ make the system highly sensitive to malicious actions, producing sharp decreases in trust scores. To quantify this sensitivity, Figure~\ref{FIG:14} (Right) presents the vulnerability score for each strategy, defined as the mean area of the trust score that falls below the safety threshold of 0.50. The results confirm that the dynamic 
$k$ setting incurs the highest vulnerability score (0.0267), reflecting its aggressive response to detected anomalies, while the fixed baselines $\alpha$ = 0.9 and
$\alpha$=0.8
yield lower scores of 0.0061 and 0.0023, respectively. This trade-off highlights a key property of the proposed model: greater sensitivity to malicious behavior necessarily results in deeper trust degradation during the attack window, which is a deliberate design choice to prevent dishonest nodes from remaining active in the consensus committee.

 The proposed dynamic trust model proves to be effective in detecting malicious activities. Its adaptability makes it particularly well-suited for interoperability scenarios, as the model can be reconfigured to accommodate varying requirements for trust degradation and recovery.

\section{Discussion}

\subsection*{Post-Quantum Cryptography Overhead}
Our evaluation demonstrates that post-quantum cryptography introduces modest end-to-end overhead, averaging 5.1\% across 30–270 TPS and decreasing to 0.2\% at peak load compared with the elliptic-curve baseline. By limiting asymmetric operations to authentication and non-repudiation while using AES-256 for bulk encryption, the proposed architecture achieves quantum-resistant interoperability with minimal performance impact.

\subsection*{Dynamic Trust Model Behavior}
The dynamic trust model's main advantage lies in detecting intermittent malicious behavior rather than convergence speed. Under sporadic attacks (epochs 40–80), dynamic $k$ reduces malicious Verifier trust to 0.001, compared with 0.1 and 0.2 for fixed-$\alpha$ baselines ($\alpha=0.9$ and $0.8$). Average trust during attacks drops to 0.506, versus 0.798 and 0.773, respectively. This stronger penalization limits malicious Verifiers' continued participation in consensus, addressing a vulnerability of fixed-weight trust models.

\subsection*{Limitations and Scope}
The evaluation is limited to two Gateways and five Verifiers ($f=1$), leaving larger-scale performance unexplored. Trust evaluation is simulation-based and does not fully capture network delays or packet loss. Additionally, the selected dynamic $k$ parameters ($15$ for penalization and $0.09$ for recovery) require further sensitivity analysis across different threat models.

\section{Conclusion}
This paper investigated cryptographic primitives for secure and efficient blockchain interoperability in notary-scheme-based architectures, evaluating lattice-based post-quantum schemes, including ML-KEM and ML-DSA, against elliptic-curve counterparts across key encapsulation, digital signatures, and cross-chain communication.

Our second contribution is a dynamic trust model that adjusts a tunable parameter $k$ to balance recent and historical behavior when evaluating node trustworthiness. Experiments under different adversarial scenarios showed that the model can adapt trust degradation and restoration to operational conditions. Compared with an EigenTrust-based baseline, our approach demonstrated improved responsiveness and accuracy, particularly in detecting malicious behavior concealed within otherwise honest actions.

Overall, the results demonstrate an adaptive trust mechanism that supports dynamic trust evaluation while maintaining security and efficiency in blockchain interoperability.

\bibliographystyle{IEEEtran}
\bibliography{refs}

@article{buterin2013ethereum,
  title={Ethereum white paper},
  author={Buterin, Vitalik and others},
  journal={GitHub repository},
  volume={1},
  pages={22--23},
  year={2013}
}

@inproceedings{hassan2025enabling,
  title={Enabling blockchain interoperability through network discovery services},
  author={Hassan, Khalid and Sokhankhosh, Amirreza and Rouhani, Sara},
  booktitle={2025 IEEE International Conference on Decentralized Applications and Infrastructures (DAPPS)},
  pages={91--100},
  year={2025},
  organization={IEEE}
}

@article{hassan2026sok,
  title={SoK: Systematizing blockchain interoperability through layered architecture and layer 2 interoperability},
  author={Hassan, Khalid and Nikraftar, Aila and Sokhankhosh, Amirreza and Esmaeilzadeh Khorasani, Koosha and Moradi, Saeed and Rouhani, Sara},
  journal={Distributed Ledger Technologies: Research and Practice},
  volume={6},
  number={1},
  pages={1--29},
  year={2026},
  publisher={ACM New York, NY}
}

@inproceedings{scheid2019bifrost,
  title={Bifr{\"o}st: a modular blockchain interoperability API},
  author={Scheid, Eder J and Hegnauer, Timo and Rodrigues, Bruno and Stiller, Burkhard},
  booktitle={2019 IEEE 44th Conference on Local Computer Networks (LCN)},
  pages={332--339},
  year={2019},
  organization={IEEE}
}

@inproceedings{karumba2023bailif,
  title={Bailif: A blockchain agnostic interoperability framework},
  author={Karumba, Samuel and Jurdak, Raja and Kanhere, Salil S and Sethuvenkatraman, Subbu},
  booktitle={2023 IEEE International Conference on Blockchain and Cryptocurrency (ICBC)},
  pages={1--9},
  year={2023},
  organization={IEEE}
}

@inproceedings{dangat2024otex,
  title={OTEx: Ownership Transfer and Execution Protocol for Blockchain Interoperability},
  author={Dangat, Shantanu and Kar, Suparna and Toshniwal, Bhavesh and Kataoka, Kotaro and Huang, Yawen and Abe, Hiroki and Xu, Xin},
  booktitle={2024 IEEE International Conference on Blockchain (Blockchain)},
  pages={165--173},
  year={2024},
  organization={IEEE}
}

@article{li2021ivycross,
  title={Ivycross: A privacy-preserving and concurrency control framework for blockchain interoperability},
  author={Li, Ming and Weng, Jian and Li, Yi and Wu, Yongdong and Weng, Jiasi and Li, Dingcheng and Xu, Guowen and Deng, Robert},
  journal={Cryptology ePrint Archive},
  year={2021}
}

@misc{axelar2021, title = {Axelar: Connecting Applications with Blockchain Ecosystems}, author = {{Axelar Team}}, year = {2021}, month = {January}, howpublished = {\url{https://www.axelar.network/whitepaper}}, note = {Accessed: 15 December 2024} }

@article{wood2016polkadot,
  title={Polkadot: Vision for a heterogeneous multi-chain framework},
  author={Wood, Gavin},
  journal={White paper},
  volume={21},
  number={2327},
  pages={4662},
  year={2016}
}

@article{belchior2021survey,
  title={A survey on blockchain interoperability: Past, present, and future trends},
  author={Belchior, Rafael and Vasconcelos, Andr{\'e} and Guerreiro, S{\'e}rgio and Correia, Miguel},
  journal={ACM Computing Surveys (CSUR)},
  volume={54},
  number={8},
  pages={1--41},
  year={2021},
  publisher={ACM New York, NY}
}

@inproceedings{dottling2023mcfly,
  title={McFly: verifiable encryption to the future made practical},
  author={D{\"o}ttling, Nico and Hanzlik, Lucjan and Magri, Bernardo and Wohnig, Stella},
  booktitle={International Conference on Financial Cryptography and Data Security},
  pages={252--269},
  year={2023},
  organization={Springer}
}

@article{madathil2022cryptographic,
  title={Cryptographic oracle-based conditional payments},
  author={Madathil, Varun and Thyagarajan, Sri AravindaKrishnan and Vasilopoulos, Dimitrios and Fournier, Lloyd and Malavolta, Giulio and Moreno-Sanchez, Pedro},
  journal={Cryptology ePrint Archive},
  year={2022}
}

@inproceedings{gerhart2024foundations,
  title={Foundations of adaptor signatures},
  author={Gerhart, Paul and Schr{\"o}der, Dominique and Soni, Pratik and Thyagarajan, Sri AravindaKrishnan},
  booktitle={Annual International Conference on the Theory and Applications of Cryptographic Techniques},
  pages={161--189},
  year={2024},
  organization={Springer}
}

@misc{loomxio,
  title = {{Intro to Loom Network}},
  author = {{Loom Team}},
  howpublished = {\url{https://loomx.io/developers/en/intro-to-loom.html}},
  note = {Accessed: March 2024},
   year={2022},
}

@misc{liquid2020,
  author = {Jonas Nick and Andrew Poelstra and Gregory Sanders},
  title = {Liquid: A Bitcoin Sidechain},
  howpublished = {\url{https://blockstream.com/assets/downloads/pdf/liquid-whitepaper.pdf}},
  year = {2020},
}

@article{alonso2020zero,
  title={Zero to monero},
  author={Alonso, Kurt M and others},
  journal={Zero to monero},
  year={2020}
}

@article{ren2023interoperability,
  title={Interoperability in blockchain: A survey},
  author={Ren, Kunpeng and Ho, Nhut-Minh and Loghin, Dumitrel and Nguyen, Thanh-Toan and Ooi, Beng Chin and Ta, Quang-Trung and Zhu, Feida},
  journal={IEEE Transactions on Knowledge and Data Engineering},
  volume={35},
  number={12},
  pages={12750--12769},
  year={2023},
  publisher={IEEE}
}

@INPROCEEDINGS{10646648,
  author={Augusto, André and Belchior, Rafael and Correia, Miguel and Vasconcelos, André and Zhang, Luyao and Hardjono, Thomas},
  booktitle={2024 IEEE Symposium on Security and Privacy (SP)}, 
  title={SoK: Security and Privacy of Blockchain Interoperability}, 
  year={2024},
  volume={},
  number={},
  pages={3840-3865},
  doi={10.1109/SP54263.2024.00255}}

@article{rivest1978method,
  title={A method for obtaining digital signatures and public-key cryptosystems},
  author={Rivest, Ronald L and Shamir, Adi and Adleman, Leonard},
  journal={Communications of the ACM},
  volume={21},
  number={2},
  pages={120--126},
  year={1978},
  publisher={ACM New York, NY, USA}
}

@article{koblitz1987elliptic,
  title={Elliptic curve cryptosystems},
  author={Koblitz, Neal},
  journal={Mathematics of computation},
  volume={48},
  number={177},
  pages={203--209},
  year={1987}
}

@article{shor1999polynomial,
  title={Polynomial-time algorithms for prime factorization and discrete logarithms on a quantum computer},
  author={Shor, Peter W},
  journal={SIAM review},
  volume={41},
  number={2},
  pages={303--332},
  year={1999},
  publisher={SIAM}
}

@inproceedings{grover1996fast,
  title={A fast quantum mechanical algorithm for database search},
  author={Grover, Lov K},
  booktitle={Proceedings of the twenty-eighth annual ACM symposium on Theory of computing},
  pages={212--219},
  year={1996}
}

@inproceedings{sharma2021analysis,
  title={Analysis of CRYSTALS-Dilithium for BlockChain security},
  author={Sharma, Lucy and Mishra, Arun},
  booktitle={2021 2nd International Conference on Secure Cyber Computing and Communications (ICSCCC)},
  pages={160--165},
  year={2021},
  organization={IEEE}
}

@inproceedings{holcomb2021pqfabric,
  title={Pqfabric: a permissioned blockchain secure from both classical and quantum attacks},
  author={Holcomb, Amelia and Pereira, Geovandro and Das, Bhargav and Mosca, Michele},
  booktitle={2021 IEEE International Conference on Blockchain and Cryptocurrency (ICBC)},
  pages={1--9},
  year={2021},
  organization={IEEE}
}

@article{wang2022secure,
  title={A secure cross-chain transaction model based on quantum multi-signature},
  author={Wang, Zhuo and Li, Jian and Chen, Xiu-Bo and Li, Chaoyang},
  journal={Quantum Information Processing},
  volume={21},
  number={8},
  pages={279},
  year={2022},
  publisher={Springer}
}

@article{yi2023post,
  title={A post-quantum blockchain notary scheme for cross-blockchain exchange},
  author={Yi, Haibo},
  journal={Computers and Electrical Engineering},
  volume={110},
  pages={108832},
  year={2023},
  publisher={Elsevier}
}

@ARTICLE{fips2024203,
  title={203: Module-Lattice-Based Key-Encapsulation Mechanism Standard},
  author={FIPS, NIST},
  journal={National Institute of Standards and Technology: Gaithersburg, MD, USA},
  year={2024}
}

@article{fips2024204,
  title={204: Module-Lattice-Based Digital Signature Standard},
  author={FIPS, NIST},
  journal={National Institute of Standards and Technology: Gaithersburg, MD, USA},
  year={2024}
}

@article{cao2023cross,
  title={Cross-chain data traceability mechanism for cross-domain access},
  author={Cao, Lifeng and Zhao, Shoucai and Gao, ZhenSheng and Du, Xuehui},
  journal={The Journal of Supercomputing},
  volume={79},
  number={5},
  pages={4944--4961},
  year={2023},
  publisher={Springer}
}

@article{chen2024cross,
  title={Cross-chain asset trading scheme for notaries based on edge cloud storage},
  author={Chen, Lang and Chen, Yuling and Tan, Chaoyue and Luo, Yun and Dou, Hui and Yang, Yuxiang},
  journal={Journal of Cloud Computing},
  volume={13},
  number={1},
  pages={90},
  year={2024},
  publisher={Springer}
}

@inproceedings{putra2023privacy,
  title={Privacy-preserving Trust Management for Blockchain-based Resource Sharing in 6G-IoT},
  author={Putra, Guntur Dharma and Dedeoglu, Volkan and Kanhere, Salil S and Jurdak, Raja},
  booktitle={2023 IEEE International Conference on Blockchain and Cryptocurrency (ICBC)},
  pages={1--9},
  year={2023},
  organization={IEEE}
}

@inproceedings{bacho2022adaptive,
  title={On the adaptive security of the threshold BLS signature scheme},
  author={Bacho, Renas and Loss, Julian},
  booktitle={Proceedings of the 2022 ACM SIGSAC Conference on Computer and Communications Security},
  pages={193--207},
  year={2022}
}

@inproceedings{victor2019measuring,
  title={Measuring ethereum-based erc20 token networks},
  author={Victor, Friedhelm and L{\"u}ders, Bianca Katharina},
  booktitle={Financial Cryptography and Data Security: 23rd International Conference, FC 2019, Frigate Bay, St. Kitts and Nevis, February 18--22, 2019, Revised Selected Papers 23},
  pages={113--129},
  year={2019},
  organization={Springer}
}

@article{tan2024study,
  title={A Study on Privacy Protection of Cross-Chain Transactions Based on Improved Notary Mechanisms},
  author={Tan, Zhong Sheng and Chen, Chun Hui and Chen, Xue Bin and Xin, Yang},
  journal={IEEE Access},
  year={2024},
  publisher={IEEE}
}

@article{wu2023distributed,
  title={A distributed cross-chain mechanism based on notary schemes and group signatures},
  author={Wu, Xiaohua and Zhang, Tingbo and Wang, Jing and Cheng, Jun and Wang, Zirui},
  journal={Journal of King Saud University-Computer and Information Sciences},
  volume={35},
  number={10},
  pages={101862},
  year={2023},
  publisher={Elsevier}
}

@inproceedings{page1999pagerank,
  title={The PageRank Citation Ranking : Bringing Order to the Web},
  author={Lawrence Page and Sergey Brin and Rajeev Motwani and Terry Winograd},
  booktitle={The Web Conference},
  year={1999},
  url={https://api.semanticscholar.org/CorpusID:1508503}
}

@inproceedings{cao2023two,
  title={Two-Phase Election Algorithm for Cross-Chain Notaries},
  author={Cao, Ling and Yang, Hao},
  booktitle={2023 3rd International Conference on Computer Science and Blockchain (CCSB)},
  pages={53--58},
  year={2023},
  organization={IEEE}
}

@article{egozcue2018variance,
  title={The variance upper bound for a mixed random variable},
  author={Egozcue, Mart{\'\i}n and Garc{\'\i}a, Luis Fuentes},
  journal={Communications in Statistics-Theory and Methods},
  volume={47},
  number={22},
  pages={5391--9395},
  year={2018},
  publisher={Taylor \& Francis}
}

@article{guo2024calculation,
  title={Calculation and selection scheme of node reputation values for notary mechanism in cross-chain},
  author={Guo, Zhihao and Hu, Xiaoming},
  journal={The Journal of Supercomputing},
  pages={1--22},
  year={2024},
  publisher={Springer}
}

@article{yang2018blockchain,
  title={Blockchain-based decentralized trust management in vehicular networks},
  author={Yang, Zhe and Yang, Kan and Lei, Lei and Zheng, Kan and Leung, Victor CM},
  journal={IEEE internet of things journal},
  volume={6},
  number={2},
  pages={1495--1505},
  year={2018},
  publisher={IEEE}
}

@article{batwa2021blockchain,
  title={Blockchain technology and trust in supply chain management: A literature review and research agenda},
  author={Batwa, Abbas and Norrman, Andreas},
  journal={Operations and Supply Chain Management: An International Journal},
  volume={14},
  number={2},
  pages={203--220},
  year={2021}
}

@article{putra2021trust,
  title={Trust-based blockchain authorization for IoT},
  author={Putra, Guntur Dharma and Dedeoglu, Volkan and Kanhere, Salil S and Jurdak, Raja and Ignjatovic, Aleksandar},
  journal={IEEE Transactions on Network and Service Management},
  volume={18},
  number={2},
  pages={1646--1658},
  year={2021},
  publisher={IEEE}
}

@article{hu2024blockchain,
  title={A Blockchain Cross-Chain Transaction Method Based on Decentralized Dynamic Reputation Value Assessment},
  author={Hu, Xiaoxuan and Ling, Yaochen and Hua, Jialin and Dong, Zhenjiang and Sun, Yanfei and Qi, Jin},
  journal={IEEE Transactions on Network and Service Management},
  year={2024},
  publisher={IEEE}
}

@inproceedings{groth2016size,
  title={On the size of pairing-based non-interactive arguments},
  author={Groth, Jens},
  booktitle={Advances in Cryptology--EUROCRYPT 2016: 35th Annual International Conference on the Theory and Applications of Cryptographic Techniques, Vienna, Austria, May 8-12, 2016, Proceedings, Part II 35},
  pages={305--326},
  year={2016},
  organization={Springer}
}

@article{petkus2019and,
  title={Why and how zk-snark works},
  author={Petkus, Maksym},
  journal={arXiv preprint arXiv:1906.07221},
  year={2019}
}

@inproceedings{kamvar2003eigentrust,
  title={The eigentrust algorithm for reputation management in p2p networks},
  author={Kamvar, Sepandar D and Schlosser, Mario T and Garcia-Molina, Hector},
  booktitle={Proceedings of the 12th international conference on World Wide Web},
  pages={640--651},
  year={2003}
}

@article{neuman1994kerberos,
  title={Kerberos: An authentication service for computer networks},
  author={Neuman, B Clifford and Ts'o, Theodore},
  journal={IEEE Communications magazine},
  volume={32},
  number={9},
  pages={33--38},
  year={1994},
  publisher={IEEE}
}

@inproceedings{zamyatin2019xclaim,
  title={Xclaim: Trustless, interoperable, cryptocurrency-backed assets},
  author={Zamyatin, Alexei and Harz, Dominik and Lind, Joshua and Panayiotou, Panayiotis and Gervais, Arthur and Knottenbelt, William},
  booktitle={2019 IEEE symposium on security and privacy (SP)},
  pages={193--210},
  year={2019},
  organization={IEEE}
}

\end{document}